\documentclass{agujournal2019}
\usepackage{color}
\usepackage{url} 
\usepackage{lineno}
\usepackage[utf8]{inputenc}

\draftfalse

\newcommand{\ssr}{    {Space Sci. Rev.}}

\newcommand{\apjl}{    {Astrophys. J. Lett.}}

\newcommand{\grl}{    {Geophys. Res. Lett.}}
\newcommand{\jgr}{    {J. Geophys. Res.}}

\usepackage{amssymb}
\usepackage{amsmath, mathtools}

\journalname{\grl}

\begin{document}

%
%



%
%




\title{Streamer-like red line diffuse auroras driven by time domain structures and ECH waves associated with a plasma injection and braking ion flows}

\authors{Yangyang Shen\affil{1}, Xu Zhang\affil{1}, Jun Liang\affil{2}, Anton Artemyev\affil{1}, Vassilis Angelopoulos\affil{1}, Emma Spanswick\affil{2}, Larry Lyons\affil{3}, Yukitoshi Nishimura\affil{4} }
\affiliation{1}{Department of Earth, Planetary, and Space Sciences, University of California, Los Angeles, CA, USA}
\affiliation{2}{Department of Physics and Astronomy, University of Calgary, Calgary, AB, Canada}
\affiliation{3}{Department of Atmospheric and Oceanic Sciences, University of California, Los Angeles, CA, USA}
\affiliation{4}{Center for Space Physics, Boston University, Boston, MA, USA}

\correspondingauthor{Yangyang Shen}{yshen@epss.ucla.edu}

\begin{keypoints}
\item THEMIS and ground-based imagers reveal streamer-like red-line auroras poleward of a black aurora and an auroral torch 
\item Red-line auroras are linked to diffuse precipitation driven by TDSs and ECH waves from an electron injection and braking ion flows 
\item Red-line auroral intensities agree with TREx-ATM forward modeling and precipitation characteristic energies agree with MSP optical inference
\end{keypoints}

\begin{abstract}

Auroral streamers are important meso-scale processes of dynamic magnetosphere-ionosphere coupling, typically studied using imagers sensitive to energetic ($>$1 keV) electron precipitation, such as all-sky imagers (ASIs). This paper reports streamer-like red-line auroras, representing low-energy ($<$1 keV) precipitation, observed poleward of a black aurora and an auroral torch. These red-line auroras were associated with a magnetospheric electron injection and braking ion flows. Observations were made using the THEMIS spacecraft and ground-based imagers, including the ASI, REGO, and meridian scanning photometer (MSP) at Fort Smith. We identify plasma sheet electron pitch-angle scattering by time-domain structures (TDSs) and electron cyclotron harmonics (ECH) waves as the driver of these red-line auroras, because of (1) a strong correlation ($\sim$0.9) between observed red-line intensities and precipitating fluxes; (2) consistent red-line intensities from auroral transport code forward modeling, and (3) consistent precipitation characteristic energies from MSP optical inference and quasi-linear estimates. 



\end{abstract}

\section*{Plain Language Summary}

Magnetic reconnection plays a key role in transporting plasma and magnetic energy in Earth's magnetosphere and ionosphere, driving space weather dynamics. The rapid plasma transport toward Earth generates plasma waves that scatter electrons, leading to electron loss into the ionosphere and the atmosphere via collisions with neutrals. In our study, we show that this wave-driven electron loss produces dynamic red auroral emissions, which are mainly caused by low-energy (less than 1,000 electron volts) electron precipitation. This wave-driven electron precipitation plays an important role in magnetosphere–ionosphere coupling and space weather modeling. For example, low-energy electrons heat the ionosphere’s F region, driving plasma upwelling and altering chemical reactions and plasma density. These changes affect ionospheric conductivities, currents, and electric potentials, as well as neutral wind dynamics in the ionosphere and thermosphere. Understanding these processes helps us better predict space weather impacts on communication, navigation, and satellite operations.

Magnetosphere transients, such as bursty bulk flows (BBFs) and plasma injections, are associated with various dynamic and localized auroral forms in the ionosphere, including poleward boundary intensifications and north-south aligned auroral streamers \cite{Angelopoulos92,Lyons99,Nakamura01,Nishimura11,Forsyth20}. Investigating the causal links between these auroral forms and magnetospheric transients improves our understanding of magnetosphere-ionosphere (MI) coupling through meso-scale processes, and helps establish high-fidelity magnetic mapping during geomagnetically active periods, such as storms and substorms, when mapping via empirical magnetospheric models is often uncertain \cite{Sergeev12,Kubyshkina19,Stephens19,Stephens23,Shen23:jgr:ELFIN_dropout,Shi24:elfin}.

Auroral streamers are well-known as the ionospheric manifestation of plasma sheet BBF channels and are linked to upward field‐aligned currents (FACs) forming at the western edge of the flow burst \cite{Nakamura01,Sergeev04:angeo,Nishimura11}. \citeA{Nakamura01} revealed that an auroral streamer is magnetically connected to only a limited part ($\leq\sim$0.4 hr magnetic local time or MLT) on the dusk flank of the entire BBF channel, which often spans $\sim$1.5 hr MLT in the dawn-dusk direction. \citeA{Gallardo-Lacourt14} found that $\sim$90\% of streamers were accompanied by enhanced equatorward flows $\sim$57 km to the east of their narrow ($<\sim$75 km) optical emissions. To date, several important questions about auroral streamers remain unresolved \cite{Forsyth20}. For example, using Polar global ultra-voilet (UV) images and conjugate DMSP observations, \citeA{Sergeev04:angeo} identified two distinct types of precipitation patterns associated with streamers. Type-I streamers exhibited monoenergetic electron precipitation from field-aligned acceleration, coinciding with narrow upward FAC regions. In contrast, Type-II streamers, often seen more equatorward and closer to diffuse auroral regions, displayed precipitation characteristics almost the same as diffuse auroras, with no evidence of field-aligned acceleration. These observations suggest that both discrete auroral processes of quasi-electrostatic coupling and diffuse auroral processes of wave-driven coupling contribute to the auroral forms associated with BBFs. It remains unclear what magnetospheric conditions or ionospheric conditions determine the occurrence of the two types of streamers. It is also unclear how streamers evolve as flow bursts or plasma ``bubbles'' decelerate in the tail-to-dipole transition region \cite{Chen&Wolf93,Dubyagin11,Yang11}, where the plasma conditions in the flow become comparable with the surrounding plasma. 

Furthermore, plasma injections are highly correlated with ion flow bursts and provide the particle source for auroral precipitation \cite{Gabrielse19}. Both braking ion flows and plasma injections are associated with various plasma waves, including whistler-mode waves \cite{Li09,Zhang18:whistlers&injections,Artemyev22:jgr:DF&ELFIN}, electron cyclotron harmonic (ECH) waves \cite{Liang11,Zhang14:ECH&DF,Zhang21:jgr:ECH}, and time-domain structures (TDSs) \cite{Malaspina15,Mozer15,Shen21:EH}. These waves collectively contribute to electron pitch-angle scattering and precipitation into diffuse auroras \cite{Meredith09:aurora,Ni16:ssr,Shen20:tds,Shen24:redline:msp,Zhang25:ECHscattering}. At $L>\sim$7 $R_E$, ECH waves and TDSs are thought to be the main drivers of the diffuse aurora \cite{Zhang15:ECH,Shen21:EH}. However, their relative contributions to meso-scale auroral structures linked to magnetospheric transients remain unclear. 

While most prior observations of streamers were obtained using whitelight, green-line (557.7 nm) or UV imagers, some flow-related auroras are better captured or can only be observed in red-line (630 nm) emissions. Red-line auroras typically result from the transition of excited atomic oxygen $O(^{1}D)$ caused by low-energy ($<$1 keV) electron precipitation above $\sim$200 km altitude \cite{Solomon88,Gillies17,Liang16:atm,Liang19:aurora}. For example, \citeA{Kepko09} used multi-spectral auroral observations to identify an equatorward-moving, flow-driven diffuse auroral patch in the red line prior to auroral expansion onset. A discrete green-line arc was observed at the westward and equatorward edge of the red line diffuse aurora. They suggested that the red-line patch was caused by enhanced pitch-angle scattering of relatively cool plasma sheet electrons. Similarly, \citeA{Liang11} and \citeA{Shen24:redline:msp} reported correlated ion flow bursts and red-line diffuse auroral patches during a geomagnetically quiet event. Combining THEMIS and Meridian Scanning Photometer (MSP) observations with forward modeling of auroras using the Trasition Region Explorer auroral transport model (TREx-ATM), they suggested that the red-line diffuse auroras were driven by pitch-angle scattering of plasma sheet electrons due to the combined effects of TDSs and whistler-mode waves. 

However, MSP observations in their studies have limited time resolution (15 s) and do not allow two-dimensional (2D) characterization of flow-driven red-line auroras. This spatiotemporal limitation hindered an exact linkage between magnetospheric wave-particle interaction processes and the observed auroras. Here using conjugate measurements between the THEMIS spacecraft and ground-based whitelight all-sky imager (ASI), Redline Emission Geospace Observatory (REGO) 2D red-line imager, and multi-wavelength MSP, all located at Fort Smith, we report streamer-like red-line diffuse auroras at the poleward edge of a black aurora and an auroral torch during the recovery phase of a substorm-like event. Our observations and model results establish a direct linkage between these red-line auroras and electron precipitation induced by TDSs and ECH waves, associated with a plasma sheet electron injection and braking ion flows. 

\section{Data and models}\label{sect2}

We use data from the THEMIS spacecraft \cite{Angelopoulos08:ssr}, equipped with the (i) Electrostatic Analyzers (ESA), which measure electron energy and pitch-angle distributions $<\sim$30 keV \cite{McFadden08:THEMIS}; (ii) Solid State Telescope (SST), which provides electron fluxes from $\sim$25 keV to $>$500 keV \cite{Angelopoulos08:sst}; (iii) Fluxgate Magnetometer, which measures DC vector magnetic field \cite{Auster08:THEMIS}; (iv) Search-Coil Magnetometer \cite{LeContel08}, Electric Field Instrument \cite{LeContel08}, and Digital Fields Board \cite{Cully08:ssr}, together providing electric and magnetic field wave power spectra with frequencies up to 4 kHz every 1 s and waveform data at 8,192 sps (DC-coupled). We also use the $SML$ index from SuperMAG \cite{Gjerloev12}.

THEMIS ASIs capture whitelight (400-700 nm) auroral images every 3 s \cite{Donovan06}. The REGO measures 2D red-line (630 nm) optical emissions every 3 s \cite{Liang16:atm}. REGO data are fully calibrated through background subtraction, flat-field correction, and Rayleigh conversion. The NORSTAR MSP measures optical luminosity in four auroral bands (470.9 nm, 486.1 nm, 557.7 nm, and 630.0 nm) \cite{Liang18:proton}. The MSP collects photons at specified wavelengths while scanning a meridian. Each high-resolution scan consists of $\sim$544 elevation steps, taking 30 s to complete a full north-to-south scan. In what follows, the ASI auroral images are typically projected to the altitude-adjusted corrected geomagnetic (AACGM) coordinates assuming an emission altitude of 110 km, and REGO and MSP red line emissions are projected assuming 230 km altitude. 

TDSs include different types of Debye-scale nonlinear electrostatic structures \cite{Mozer15}, and electron phase space holes are the predominant type in the near-Earth plasma sheet \cite{Ergun15, Malaspina18}. Electron scattering by electron holes can be quantified using the quasi-linear approach initially developed for an ensemble of plane waves \cite{Kennel&Engelmann66}, assuming statistically uniform and weak TDS turbulence and electron random gyrophase between interactions \cite{Vasko17:diffusion,Vasko18:pop}. We use the local diffusion coefficients $D_{\alpha\alpha}$ derived in \citeA{Vasko18:pop} and refined by \citeA{Shen21:EH}, incorporating statistical TDS properties such as parallel scales and propagating velocities. Importantly, the local $D_{\alpha\alpha}$ does not depend on individual TDS electric fields but on the root‐mean‐square intensity $E_w$ derived from spectra data, which take into account the occurrence of TDSs \cite{Vasko18:pop,Shen21:EH}.  

The local $D_{\alpha\alpha}$ due to ECH waves can be computed using the equations derived in \citeA{Lyons74} and \citeA{Horne&Thorne00}. ECH waves are likely generated by loss-cone distributions with wave normal angles $\theta_{kB}>\sim$85$^\circ$ \cite{AshourAbdalla79} or beam distributions with $\theta_{kB}<$85$^\circ$ \cite{Zhang21:jgr:ECH,Zhang25:ECHscattering}. Because ECH waves have wave vector $k$ parallel to wave $\vec{E_{w}}$, we can calculate the ECH wave normal angle from the ratio of parallel to total wave electric fields $\theta_{kB}$= $\arccos{E_{||}/E_{tot}}$ \cite{Zhang21:jgr:ECH}. To calculate $k$, we fit the observed electron distribution with either a loss-cone distribution or a beam distribution and use Waves in Homogeneous Anisotropic Magnetized Plasma (WHAMP) to solve the dispersion relation \cite{Zhang21:jgr:ECH,Zhang25:ECHscattering}. The dispersion relation is then used to calculate the local $D_{\alpha\alpha}$.

Local $D_{\alpha\alpha}$ can be bounce-averaged by the standard procedure as \cite{Glauert&Horne05}: $\langle D_{\alpha_{eq} \alpha_{eq} }\rangle  = v^{-1} \tau^{-1} \int {D_{\alpha\alpha} \left( {\partial \alpha_{eq} /\partial \alpha } \right)^2 \left( {\partial s /\partial \lambda} \right)d\lambda /\cos \alpha}$, where the integration is over the period of bounce motion $\tau_B$, and $s=s(\lambda)$ is the length of a field line. Because ECH waves are confined near the magnetic equator \cite{Meredith09:aurora}, the maximum latitude where waves occur is assumed to be $\pm$3$^\circ$. The latitudinal distribution of wave intensity is obtained from \citeA{Zhang21:jgr:ECH} and is then normalized to the observed wave intensities by THEMIS. Because TDS-driven $D_{\alpha\alpha}$ are most effective near the equator than at higher latitudes, we assume the maximum latitude to be $\pm$25$^\circ$ \cite{Shen21:EH,Shen24:redline:msp}. We use the field line electron density model from \citeA{Denton06}, normalized to the observed local densities near the equator. We use empirical Tsyganenko T01 magnetic field models for bounce-averaging \cite{Tsyganenko02}. We adjust latitudinal profiles of the magnetic field intensity near the equator using spline function fits to local measurements of THEMIS \cite{Shen24:redline:msp}.

Assuming isotropy within the loss cone and quasi-steady equilibrium \cite{Ni12:ECH}, the differential precipitating energy flux within the loss cone can be estimated as $x(E){j\left( {E,\alpha_{LC} } \right)}$, where 
$x(E)=2\int_{0}^{1}I_{0}(Z_{0}\tau)\tau d\tau/ {I_{0}(Z_{0})}$, being the index of loss cone filling, $j(E,\alpha_{LC})$ is the differential energy flux at the edge of the loss cone, $I_0$ is the modified Bessel function with an argument $Z_0\simeq\alpha_{LC}/\sqrt{ D_{\alpha\alpha,LC}\cdot\tau_{loss}}$ \cite{Kennel&Petschek66}, $D_{\alpha\alpha,LC}$ is the bounce-averaged diffusion coefficient near the loss cone, and $\tau_{loss}$ is assumed to be half of the bounce period. The precipitating distributions due to individual waves are obtained from wave-specific $D_{\alpha\alpha,LC}$, whereas the total precipitating distributions are determined by the sum of $D_{\alpha\alpha,LC}$ due to combined scattering by all wave modes. We perform Maxwellian-type fits to each precipitation energy distribution to estimate the characteristic energy \cite{Ni14}.

The TREx-ATM simulates the transport of auroral electrons in the upper atmosphere, including ionization, heating, secondary electron production, and impact excitation of neutrals \cite{Liang16:atm,Liang24}. The model computes the photon volume emission rate at various wavelengths based on inputs of full electron precipitation distributions. We include ambipolar diffusion, interhemispheric coupling of backscattered secondaries \cite{Khazanov21,Solomon20}, and dominant chemical processes in the E and F region above 80 km, where Pederson and Hall conductivity mainly resides \cite{Liang22:model_said}. Notably, the time-dependent and self-consistent inclusion of plasma heating and ambipolar diffusion is crucial for proper modeling red-line emissions from soft electron precipitation. This is because the red-line emission is known to have a long radiative timescale ($\sim$110 s) and may include non-trivial contributions from thermal emissions in the F-region ionosphere \cite{Solomon88}. We adopt the NRLMSISE-2000 model \cite{Picone02} for the ambient neutral atmosphere and IRI2016 \cite{Bilitza17} for the background ionosphere parameters corresponding to the times and locations of our observations.

\section{Results}

\begin{figure}
\centering
\hspace*{-1.5cm}
\includegraphics[scale=0.98,angle=0]{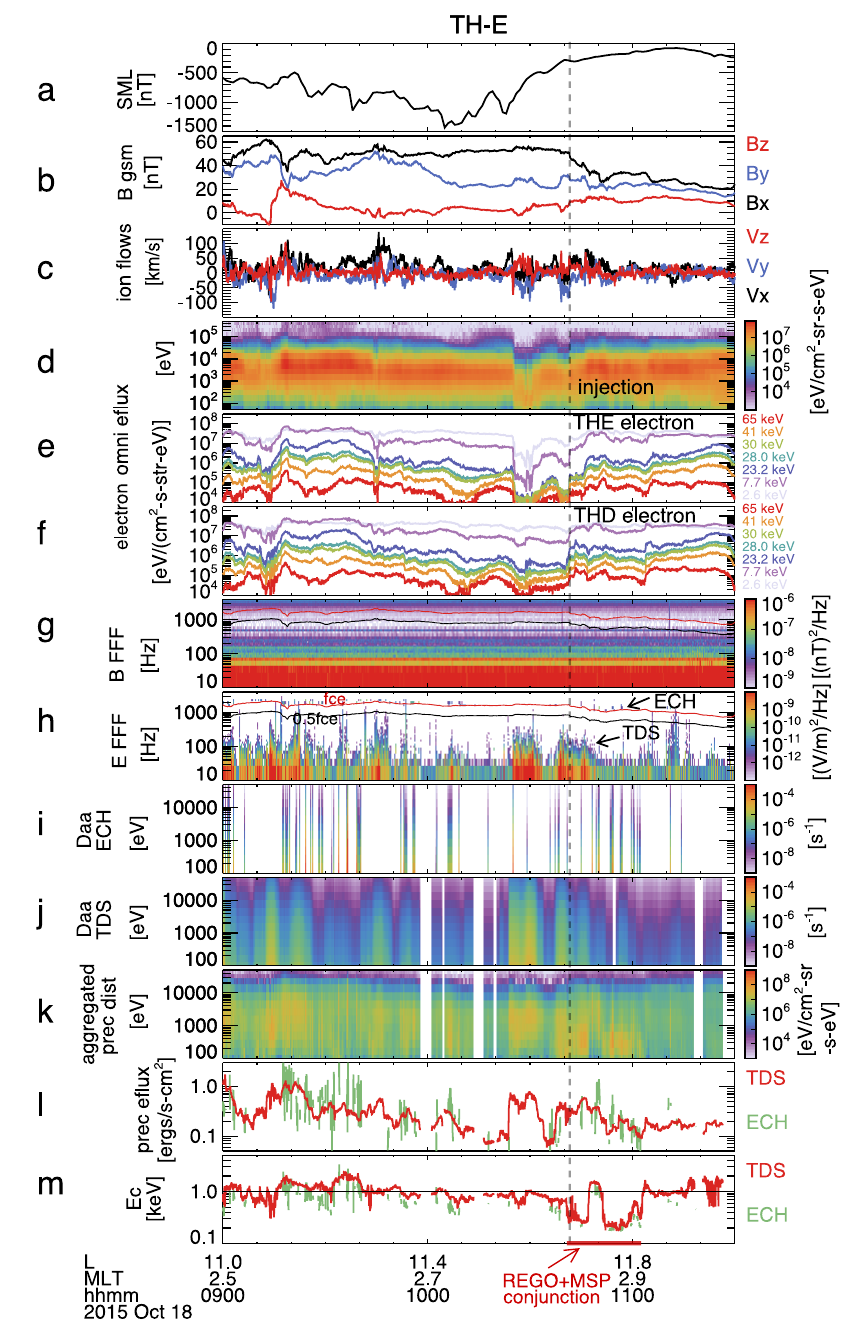}
\caption{(a) $SML$ index. (b) THEMIS-E magnetic field. (c) Ion flow velocity. (d) Electron energy-time spectrogram. (e-f) THEMIS-E and THEMIS-D electron energy flux in 2.6--65 keV. (g) wave magnetic field spectra. (h) wave electric field spectra. (i) ECH-driven electron bounce-averaged diffusion coefficients $\langle D_{\alpha\alpha}\rangle$. (j) TDS-driven $\langle D_{\alpha\alpha}\rangle$. (k) Aggregated precipitating electron distributions. (l) Integral precipitating electron energy fluxes. (m) Precipitation electron characteristic energy $E_c$. }
\vspace*{-1.0cm}
\label{fig1}
\end{figure}

The substorm-like event occurred between $\sim$09:00--11:30 UT on Oct 18, 2015, during a moderate geomagnetic storm as $D_{st}$ descended to -56 nT. The expansion and recovery phases of the substorm-like event were identified near 08:46 UT and 10:05 UT, respectively, without a distinct growth phase \cite{Forsyth15}. The $SML$ index decreased to $\sim$-1500 nT near 10:05 UT(Fig.~\ref{fig1}a). During the recovery phase near 10:40 UT, the THEMIS-E spacecraft, positioned in the tail-to-dipole transition region with $L\sim$11.8 and $MLT\sim$2.9 hr, observed an electron injection with energy flux enhancements across a broad range of energies up to $\sim$200 keV, along with weak but noticeable earthward ion flows with $V_x$ exceeding 50 km/s (Fig.~\ref{fig1}c--\ref{fig1}f). This injection was detected by both THEMIS-E and THEMIS-D near the equator. Note that THEMIS-E briefly traversed the plasma sheet boundary layer between 10:25--10:40 UT before detecting the injection (Fig.~\ref{fig1}d). During the event, the plasma sheet electron temperature fluctuated between $\sim$1--2 keV. 

Fig.~\ref{fig1}b shows a gradual increase in the magnetic field $B_z$ by $\sim$10 nT and a decrease in $B_x$ by $\sim$30 nT, indicating that the spacecraft was slowly moving toward the central plasma sheet. In contrast, the observed injection flux increase was more abrupt, distinct from mapping changes associated with magnetic field variations. By estimating the dependence of flux variations on magnetic field $B_x$ (see Supporting Information or SI), we adjusted THEMIS measured energy fluxes to obtain statistically medium and maximum equatorial flux levels while preserving the energy distribution \cite{Shen24:redline:msp}. These position-independent, corrected equatorial energy flux data were then used to estimate precipitating electron distributions in subsequent analyses. The equatorial flux adjustments assume that variations in precipitating energy flux are mainly driven by variations in wave properties \cite{Nishimura10:Science,Kasahara18:nature}, while equatorial particle flux levels remain relatively stable during the period of interest (Fig.~\ref{fig1}f).

Fig.~\ref{fig1}g--\ref{fig1}h show that broadband electrostatic fluctuations persisted below $\sim$1 kHz, ECH waves above $f_{ce}$ were intermittent around the injection, and no whistler-mode waves were detected throughout the event. Although waveform data were unavailable during 10:00--11:30 UT, those available around the injection event near 09:00-09:20 UT revealed that these electrostatic broadband fluctuations consisted of TDSs at frequencies above $\sim$50 Hz \cite{Shen21:EH}. Examples of waveform measurements showing TDSs and ECH waves are provided in SI. We estimate TDS $E_w$ by integrating electric field power within $\sim$30--900 Hz and ECH $E_w$ within $f_{ce}$--2$f_{ce}$ and 2$f_{ce}$--3$f_{ce}$, including the first and second harmonics. Wave normal angles of ECH waves were predominantly $>$85$^\circ$, consistent with loss-cone driven instabilities \cite{Zhang21:jgr:ECH}.  

Using in-situ plasma and field measurements, we calculate the bounce-averaged $D_{\alpha\alpha}$ for TDSs and ECH waves (see Fig.~\ref{fig1}i--\ref{fig1}j and Section~\ref{sect2}). These time-varying $D_{\alpha\alpha}$, combined with measured equatorial electron energy fluxes near the loss cone, are used to estimate the aggregated and constituent precipitating electron distributions within the loss cone, resulting from electron scattering by combined and individual wave effects (Fig.~\ref{fig1}k--\ref{fig1}l). Fig.~\ref{fig1}l presents integral precipitating energy fluxes in the range of $\sim$100 eV to 50 keV, which exhibit fluctuations strongly dependent on the wave properties, such as TDS and ECH $E_w$. The precipitating energy fluxes vary between 0.1--2.0 ergs/cm$^2$-s, indicating weak to moderate diffuse auroral emissions \cite{Newell09,Ni12:ECH}. Note that the precipitating fluxes in Fig.~\ref{fig1}l are likely underestimated by a factor of $\sim$2--3 during intense flux periods, because medium equatorial flux levels were used instead of the maximum values. Precipitating fluxes driven by TDS activities persisted throughout the event but were generally weaker than those driven by ECH waves, which were otherwise more intermittent and bursty. The characteristic energies of the precipitating distributions driven by TDSs and ECH waves were comparable, remaining near 1 keV for most of the time (Fig.~\ref{fig1}m). This is because both TDSs and ECH waves primarily scatter electrons with energies less than a few keV \cite{Zhang15:ECH,Shen21:EH,Zhang25:ECHscattering}. 

\begin{figure}
\centering
\hspace*{-1.5cm}
\includegraphics[scale=0.85,angle=0]{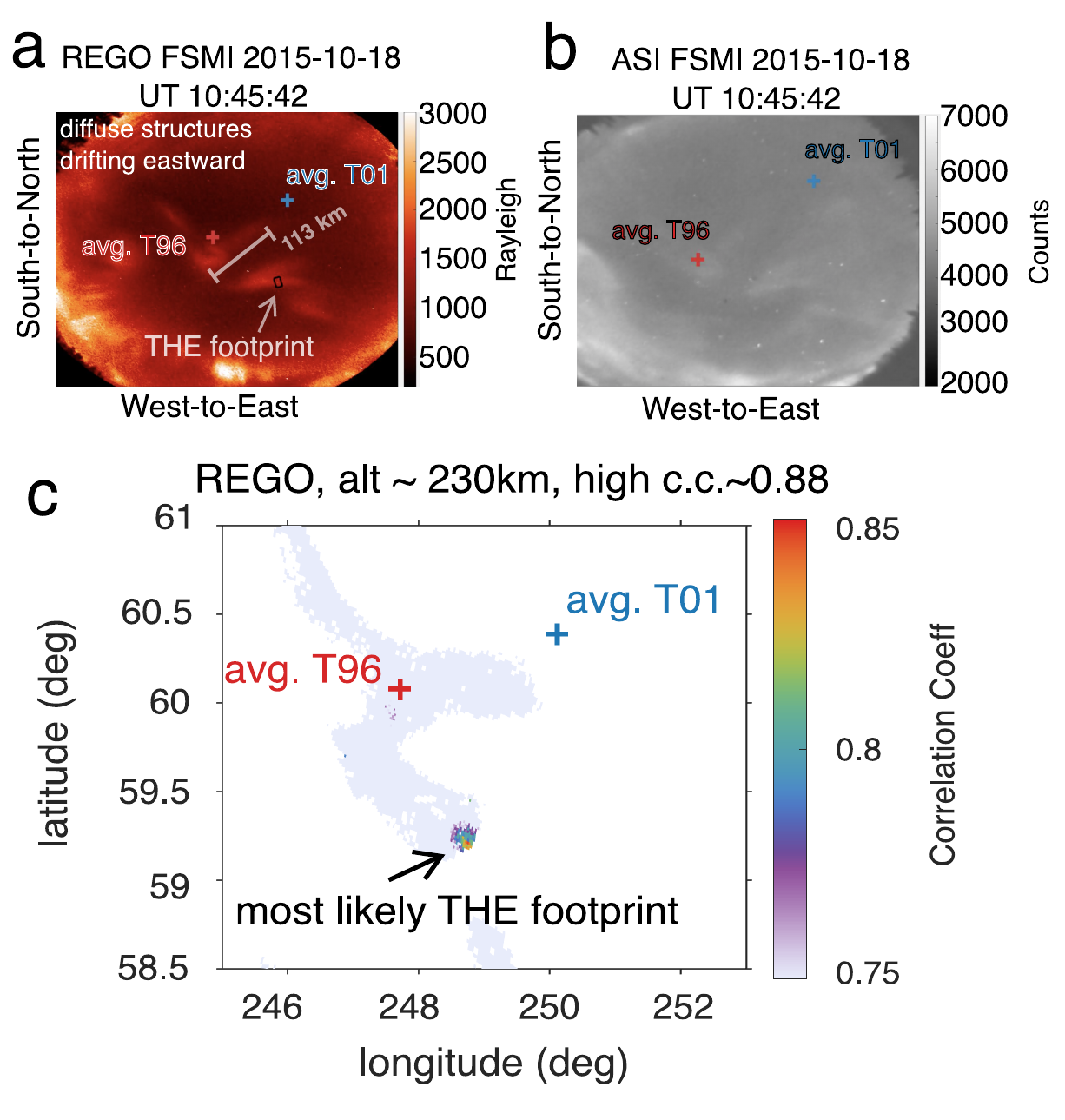}
\caption{(a) Example REGO red-line auroral image showing streamer-like diffuse auroral structures with horizontal spatial scales of up to $\sim$100 km drifting eastward and equatorward, recorded at 10:45:42 UT. (b) Example ASI whitelight auroral image corresponding to the red-line image. The streamer-like structures are less clear, blended into the diffuse auroral background. (c) Cross-correlation coefficients between red-line auroral intensities measured by the REGO 2D imager and TDS-driven precipitating electron energy fluxes estimated from THEMIS-E. The maximum correlation reached $\sim$0.88 around the geographic latitude of 59.1$^\circ$ and longitude of 248.9$^\circ$, which we interpret as the most likely THEMIS-E footprint during the 20-min interval from 10:44--11:04 UT. The average footprints of THEMIS-E predicted by the T01 model are also indicated by the plus symbol.  }
\vspace*{-1.0cm}
\label{fig2}
\end{figure}

Following the injection near 10:40 UT, precipitation characteristic energies decreased to below 0.5 keV during two intervals between 10:40--11:05 UT (Fig.~\ref{fig1}m). Spanning these periods, the THEMIS-E spacecraft was approximately conjugate with eastward- and equatorward-drifting streamer-like diffuse auroral filaments located at the poleward edge of a black aurora and an auroral torch. Movies from the ASI and REGO detailing the auroral dynamics have been provided in SI. Fig.~\ref{fig2}a-\ref{fig2}b display examples of REGO and ASI images highlighting streamer-like diffuse auroral structures. These structures spanned spatial scales of several to $\sim$100 km when mapped to 230 km altitude. The streamer-like auroras appeared patchy and were embedded within a uniform diffuse background in ASI images, with no evidence of field-aligned acceleration as often seen in Alfv{\'e}nic ray-like or vortex-like structures \cite{Lynch12,Liang19:aurora,Kataoka21}. Furthermore, their variations occurred on time and spatial scales smaller than, thus distinct from, those of field-line resonance arcs reported by \cite{Gillies18:flr}.  


To find the linkage between magnetospheric observations and ionosphere auroras, we adopt the method of \citeA{Nishimura10:Science}, which identified the driver of pulsating auroras by correlating the intensity modulation of lower‐band chorus with quasi-periodic pulsating auroral emissions near the spacecraft’s magnetic footprint. In our study, the temporal correlation between waves and auroras is enabled by the drifting, patchy nature of the auroral structures traversing the spacecraft's ionospheric footprints. We evaluate the temporal cross-correlation between the auroral intensity variations---for both red-line and whitelight 2D images---and variations in magnetospheric precipitating electron energy flux. During 10:44–11:04 UT, we obtain the max correlation coefficients based on the following criteria: (i) lag times for correlation are restricted to $<$6 s (within 2 image frames); (ii) emission altitudes are assumed 230 km for REGO and 110 km for the ASI; (iii) elevation angles of image pixels for analysis are limited to $>$45$^\circ$. 

Fig.~\ref{fig2}c shows that the maximum correlation coefficients between TDS-driven precipitating energy fluxes and REGO red-line auroral intensities reached $\sim$0.9 within a spatial region spanning $\sim$0.2$^{\circ}\times$0.3$^\circ$ centered at a geographic latitude of 59.1$^\circ$ and longitude of 248.9$^\circ$. This strong correlation indicates that the average THEMIS-E footprint during the 20-min interval was likely near this position. In comparison, the average footprints of THEMIS-E predicted by the Tsyganenko T96 and T01 models were offset by $\sim$1$^\circ$ latitude poleward and $\sim$2$^\circ$ longitude. In addition, the correlation between TDS-driven precipitation and whitelight auroral intensities remained below 0.7, while the correlations between ECH-driven precipitation and auroral intensities were even weaker.

\begin{figure}
\centering
\hspace*{-1.5cm}
\includegraphics[scale=1.0,angle=0]{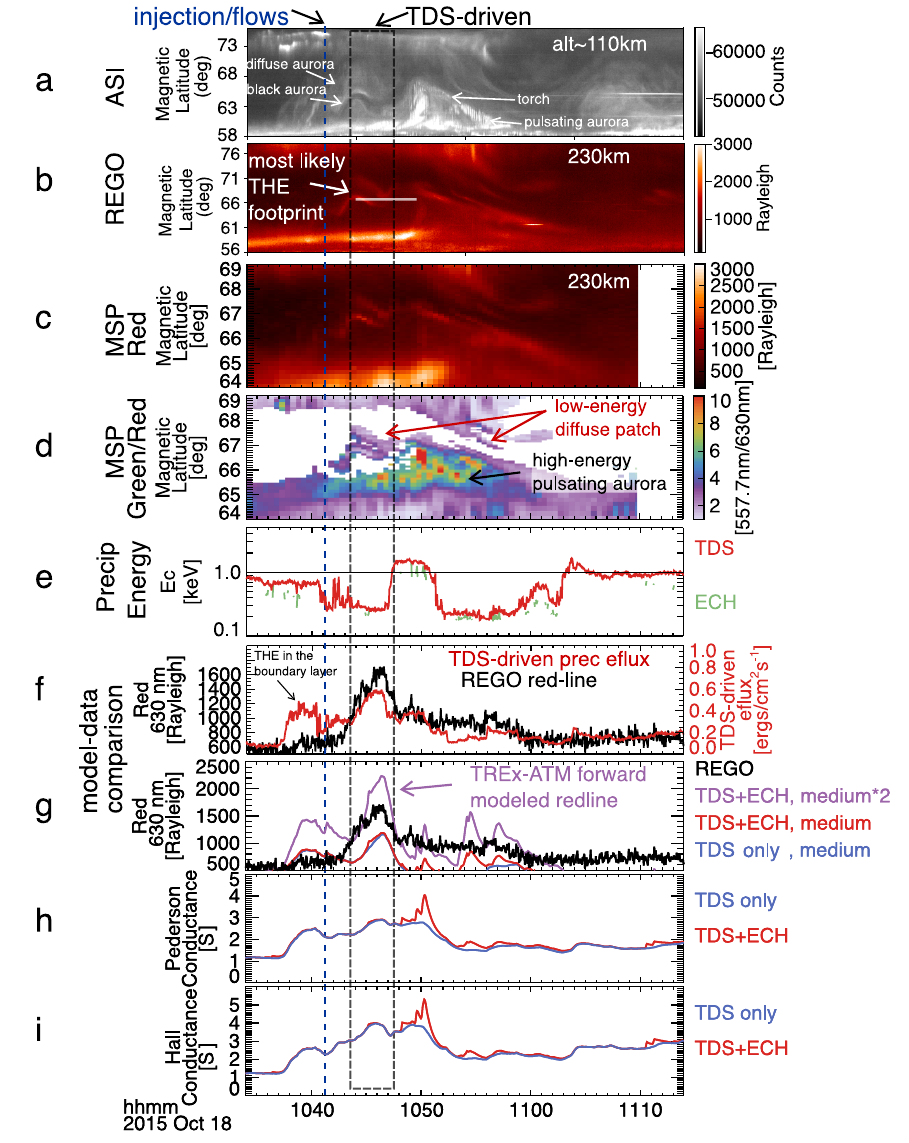}
\caption{(a) THEMIS ASI auroral keogram. (b) REGO red-line auroral keogram. (c) MSP red-line emission keogram. (d) MSP green-to-red (557.7-to-630 nm) emission ratio keogram. (e) Precipitating electron characteristic energies, same as that in Fig.~\ref{fig1}m. (f) Comparison between REGO red-line auroral intensities (black) measured at the most likely THEMIS-E footprint and TDS-driven precipitating energy fluxes (red). (g) Comparison between REGO red-line auroral intensities (black) and TREx-ATM forward modeled auroral red-line intensities. The red and magenta lines represent forward-modeled intensities resulting from the combined effects of TDS and ECH waves, using equatorial medium flux levels and twice the medium flux levels, respectively. The blue line shows forward-modeled intensities due to TDS scattering using medium flux levels. (h) Pederson conductance due to TDS-driven alone (blue) and combined TDS and ECH wave driven (red) diffuse auroral precipitation. (i) Hall conductance in the same format as panel (h).  }
\label{fig3}
\end{figure}

Fig.~\ref{fig3} provides a detailed comparison of auroral features observed by ASI, REGO, and MSP between 10:34--11:04 UT, along with estimated precipitating energy fluxes and characteristic energies from THEMIS-E measurements, and TREx-ATM forward modeled auroral red-line intensities and ionosphere conductance variations. The ASI auroral keogram in Fig.~\ref{fig3}a displays a large region of diffuse auroras below $\sim$70$^\circ$ magnetic latitude, embedding a black aurora and an auroral torch harboring pulsating auroras---typical features of the substorm recovery phase \cite{Akasofu64,Henderson12}. While differences in imager field-of-view, elevation angles, and assumed emission heights may cause significant discrepancies in the projected magnetic latitudes of auroras observed by ASI, REGO, and MSP, the correspondence between auroral features across the three datasets is clearly identifiable. For example, all three imagers captured the $\Omega$-shaped auroral enhancement at the poleward edge of the black aurora, which occurred immediately after THEMIS-E detected the injection and braking ion flows near 10:40 UT. The trailing poleward edge of the auroral torch is also well defined in all three imagers (Fig.~\ref{fig3}a--\ref{fig3}c). 


Distinct features of eastward- and equatorward-drifting streamer-like auroras were observed near 10:45 and 10:55 UT, located just poleward of the black aurora and the auroral torch. These streamer-like features were clearly visible in the red-line emission but seamlessly veiled into the diffuse background in the ASI whitelight emission. Using a lookup table constructed by TREx-ATM modeling, the measured MSP green-to-red (557.7-to-630 nm) emission ratios indicate that precipitating electron characteristic energies associated with these streamer-like red-line auroras were $\sim$0.2--0.5 keV (Fig.~\ref{fig3}d). This is consistent with $\sim$0.3 keV precipitation characteristic energies estimated from THEMIS-E, resulting from TDS-induced plasma sheet electron pitch-angle scattering (Fig.~\ref{fig3}e). During 10:48--10:51 UT, as the magnetic footprints of THEMIS-E traversed the poleward edge of the auroral torch, the precipitation characteristic energies increased to above 1 keV. This increase correlated with the increase in MSP green-to-red emission ratio and corresponding characteristic energies near the auroral torch and pulsating auroras. It is likely that ECH waves contributed significantly to these high-energy auroral precipitation (Fig.~\ref{fig1}l and Fig.~\ref{fig3}e, \cite{Liang16:atm}). However, understanding the drivers of these higher-energy auroral features associated with the auroral torch and pulsating auroras is beyond the scope of this study. 

Fig.~\ref{fig3}f demonstrates a strong correlation between REGO red-line auroral intensity variations and TDS-driven precipitating energy fluxes, suggesting that TDSs are likely the driver of these meso-scale auroral forms. The red-line intensity enhancements correspond to streamer-like structures near the THEMIS-E footprint, which is inferred from the correlation analysis presented in Fig.~\ref{fig2}c. To evaluate whether TDS-driven precipitating electron distributions can reproduce the observed auroral intensity variations, we use the TREx-ATM auroral transport code to forward model red-line auroras to compare them with observations (Fig.~\ref{fig3}g). This model-observation comparison suggests that TDS-driven precipitation adequately explains the red-line intensity enhancements of the streamer-like structures near 10:45 UT. However, ECH-driven precipitation played a dominant role in the red-line auroral structures near 10:50 UT and between 10:54--10:58 UT. It is important to note that these red-line emissions have included significant contributions from thermal emissions, caused by elevated electron temperatures resulting from the low-energy diffuse auroral precipitation in the F-region ionosphere. Furthermore, Fig.~\ref{fig3}h--\ref{fig3}i show that the diffuse auroral precipitation associated with TDSs and ECH waves can enhance both Pederson and Hall conductance by more than 4 S in the nightside auroral ionosphere.


\section{Discussion}

Most prior studies on auroral streamers have used whitelight, green-line, or UV imagers, which primarily respond to energetic ($>$1 keV) electron precipitation. These energetic streamers are often driven by quasi-electrostatic MI coupling via intense upward FACs and monoenergetic electron precipitation \cite{Lyons99,Nakamura01,Sergeev04:angeo,Nishimura11,Yang11,Forsyth20}. However, a few studies have shown that certain types of streamers are associated with diffuse auroral processes, where plasma sheet electrons are pitch-angle scattered by waves embedded within BBF channels and plasma injections. For example, the flow-driven red-line diffuse auroral patch reported by \citeA{Kepko09} was attributed to enhanced pitch-angle scattering of cool plasma sheet electrons. \citeA{Sergeev04:angeo} revealed that Type-II streamers near the diffuse auroral oval exhibited characteristics of non-accelerated diffuse auroral precipitation. In the nightside tail-to-dipole transition region, braking BBFs and plasma injections are known to generate various plasma waves, including ECH waves, TDSs, and whistler-mode waves \cite{Li09,Zhang14:ECH&DF,Mozer15,Shen21:EH,Artemyev22:jgr:DF&ELFIN}. When flow-related electrostatic coupling is likely weakened---due to reduced contrast between bubble structures and the background plasma---wave-driven coupling via diffuse auroral processes may instead take control, provided that the plasma sheet conditions are prime for diffuse auroral precipitation. The results in our paper, together with recent findings from \citeA{Shen24:redline:msp}, provide evidence that magnetospheric electron injections and braking ion flows couple to ionosphere streamer-like red-line auroras via diffuse auroral precipitation.

The streamer-like red-line auroras observed in our event were located at the poleward edge of a black aurora and an auroral torch during the substorm recovery phase (Fig.~\ref{fig3}a--\ref{fig3}d). Recently, \citeA{Spanswick24:natcom} have presented numerous observations of continuum emissions intimately coupled to meso-scale auroral dynamics, such as aurora torches, using TREx broadband color auroral imager (TREx-RGB). In their main event, clear red-line auroral structures were observed at the poleward edge of an auroral torch, similar to those in our event. Additionally, observations from the LAMP rocket mission revealed structured red-line emissions adjacent to a black aurora and an auroral torch entraining pulsating auroras during a substorm recovery phase \cite{Hosokawa24,Lessard24}. The LAMP low-energy electron measurements showed that the precipitating electron distributions corresponding to the red-line auroras had energies of only a few hundred eV with isotropic pitch-angle distributions, indicative of pitch-angle scattering of plasma sheet electrons. Thus, the red-line auroral features reported in our event are not uncommon and may represent a unique category of meso-scale auroral forms.

\section{Conclusion}

Combining conjugate observations between THEMIS spacecraft and ground-based auroral imagers, including ASI, REGO, and MSP, we report unique streamer-like red-line auroral structures located at the poleward edge of a black aurora and an auroral torch during the recovery phase of a substorm-like event. The streamer-like auroras were associated with an electron injection and braking ion flows in the nightside tail-to-dipole transition region. By correlating the observed red-line auroral intensities and precipitating fluxes driven by electron pitch-angle scattering from TDSs and ECH waves, we establish a direct linkage between the streamer-like red-line auroras and wave-induced diffuse auroral precipitation from the electron injection. Multi-wavelength auroral emission ratios from MSP indicate precipitation characteristic energies of $\sim$0.2--0.5 keV, consistent with those of quasi-linear estimates from THEMIS measurements. Auroral forward modeling using TREx-ATM further supports that these red-line auroras can be attributed to precipitation induced by TDSs and ECH waves.   

These streamer-like red-line diffuse auroras represent a significant alternative pathway for MI coupling, involving injections, braking ion flows, and wave-driven diffuse auroral processes, distinct from the conventional paradigm of streamer quasi-electrostatic coupling via field-aligned acceleration and monoenergetic electron precipitation. Our findings also demonstrate the feasibility of using REGO red-line imagers to complement whitelight or greenline ASI imagers for studying low-energy auroral streamers, and for remote sensing meso-scale transients and the associated kinetic processes in the tail-to-dipole transition region.

\acknowledgments
Y.Shen appreciates helpful discussions with Marc Lessard on LAMP Rocket observations. This work has been supported by NASA projects 80NSSC23K0413 and 80NSSC23K0108. X. Zhang is supported by NASA grant 80NSSC22K1638. TREx-ATM is supported by Canadian Space Agency. We acknowledge the support of NASA contract NAS5-02099 for the use of data from the THEMIS Mission.

\section*{Open Research} \noindent 

THEMIS data is available at \url{http://themis.ssl.berkeley.edu/data/themis/}. REGO and THEMIS ASI data are publicly available from \url{http://themis.ssl.berkeley.edu/data/themis/thg/l2/asi}, and NORSTAR MSP are accessible from \url{https://data.phys.ucalgary.ca/} sorted by project/GO-Canada/GO-Storm/msp/. TREx-ATM is accessible via API or PyAuroraX at \url{https://data.phys.ucalgary.ca/} under working with data. Data access and processing was done using SPEDAS V4.1, see \citeA{Angelopoulos19}.



%
%

%
%
%
%
%

\end{document}


%
%


\title{Supporting Information for ``Streamer-like red line diffuse auroras driven by time domain structures and ECH waves associated with a plasma injection and braking ion flows''}

\authors{Yangyang Shen\affil{1}, Xu Zhang\affil{1}, Jun Liang\affil{2}, Anton Artemyev\affil{1}, Vassilis Angelopoulos\affil{1}, Emma Spanswick\affil{2}, Larry Lyons\affil{3}, Yukitoshi Nishimura\affil{4} }
\affiliation{1}{Department of Earth, Planetary, and Space Sciences, University of California, Los Angeles, CA, USA}
\affiliation{2}{Department of Physics and Astronomy, University of Calgary, Calgary, AB, Canada}
\affiliation{3}{Department of Atmospheric and Oceanic Sciences, University of California, Los Angeles, CA, USA}
\affiliation{4}{Center for Space Physics, Boston University, Boston, MA, USA}

%


%
%

%

\begin{article}

%
%

\noindent\textbf{Contents of this file}
\begin{enumerate}
\item Figure S1 
\item Figure S2
\item Figure S3
\item Movie S1
\item Movie S2
\item Movie S3
\end{enumerate}

\noindent\textbf{Introduction} \\

This Supporting Information includes Figure S1 presenting the Fort Smith all-sky imager (ASI) keogram of the auroral activity occurring between 09:30--11:30 UT on October 18, 2015. The auroral features of our interest in the main text spanned 10:34--11:04 UT, including the broad region of the diffuse auroras, a black aurora, and an auroral torch. Corresponding to this keogram, we have included Movie S1 to show the original THEMIS ASI auroral images for the event. In addition, we have included Movie S2 and S3 to show geographically mapped auroral observations from ASI (assuming 110 km altitude) and REGO (assuming 230 km altitude) between 10:44--11:04 UT, when meso-scale streamer-like red-line auroras were observed.

Figure S2 illustrates the dependence of parallel electron energy fluxes (0$^\circ$--22.5$^\circ$ in pitch angle) near the loss cone on the magnetic field $B_x$, indicative of the spacecraft's proximity to the true magnetic equator. The observed flux variability for a fixed energy channel primarily results from changes in the magnetic field configuration, except for the injection period near 10:40 UT. To accurately calculate precipitating electron distributions corresponding to ground-based red-line diffuse auroral observations, it is necessary to use realistic equatorial fluxes measured when $|B_x|$ is near its minimum at the equator and the spacecraft is located in the central plasma sheet. We select equatorial flux data from THEMIS-D and THEMIS-E during intervals when $|B_{x}|$ is small $<$35 nT and equatorial fluxes stabilize within this $|B_{x}|$ range. The full survey mode ESA and SST electron energy flux data were collected between 09:30--11:30 UT. Panel c of Figure S2 demonstrates the equatorial energy flux spectra, showing the statistical medium (black dashed line) and maximum (black solid line) values of measured fluxes for each energy channel. In calculating precipitating electron distributions, we keep THEMIS-measured local energy distribution unchanged but proportionally adjust the maximum flux level (if it is smaller) to the statistical medium equatorial fluxes according to Panel c. The flux-adjusted energy distributions are used to estimate precipitating electron distributions and to drive the TREx-ATM model runs.

We have also included one figure (Figure S3) that displays example waveform measurements of time domain structures (TDSs) and electron cyclotron harmonic (ECH) waves observed between 09:00--09:20 UT around an earlier injection event. Panels a and b show that TDSs have frequencies ranging from $\sim$50 Hz up to 3 kHz in wavelet spectrogram, and the solitary TDS or electron phase space hole structures can be identified by the millisecond-scale bipolar parallel electric fields. Panels c and d show that both the first and second harmonics of ECH waves were observed, with wave fluctuations mainly in the perpendicular $E_y$ component and wave normal angles $>$85$^\circ$.

\newpage

\end{article}


%
%
%
%

\begin{figure}
\setfigurenum{S1}
\centering
\includegraphics[scale=0.45,angle=0,origin=c]{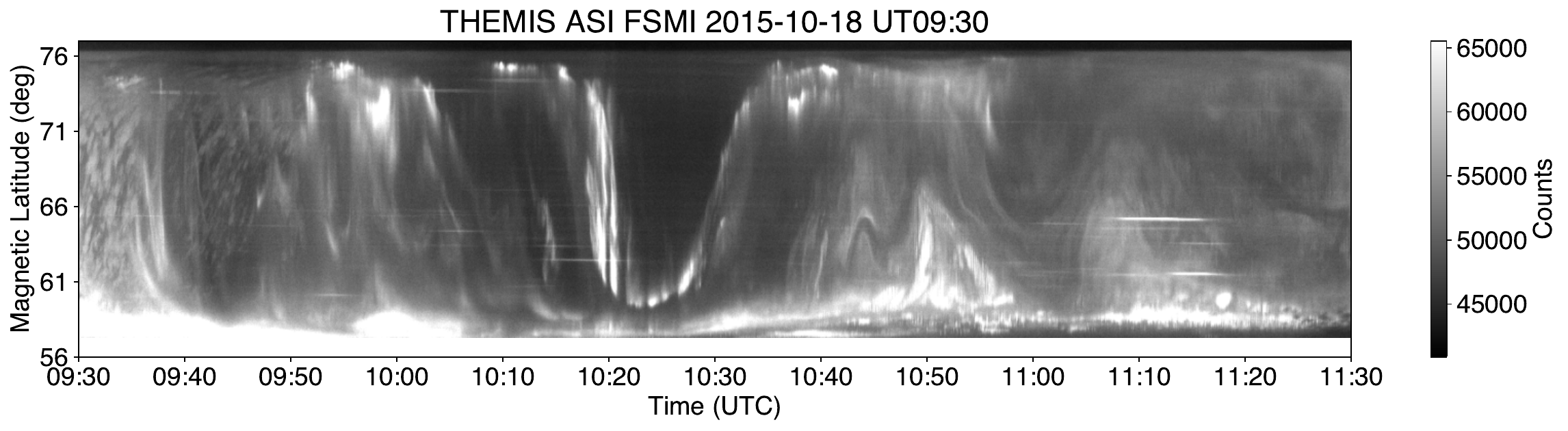}
\end{figure}
\newpage
\begin{figure}
\caption{THEMIS ASI auroral keogram showing an overview of the auroral features during the substorm-like event between 09:30--11:30 UT. }
\label{figS1}
\end{figure}
\clearpage
\newpage

\begin{figure}
\setfigurenum{S1}
\centering
\includegraphics[scale=0.37,angle=0,origin=c]{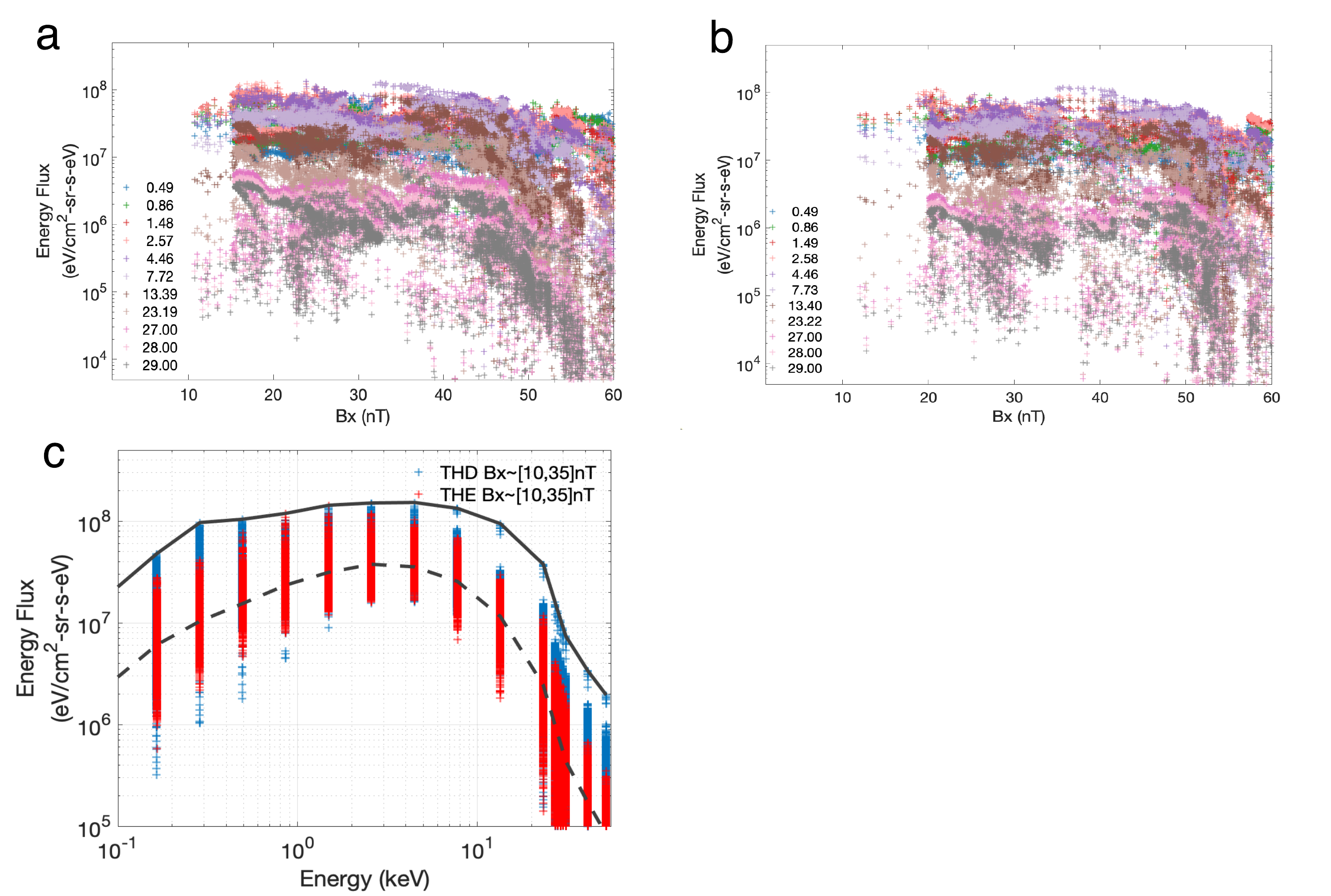}
\end{figure}
\newpage
\begin{figure}
\caption{(a) THEMIS-D measured parallel electron energy flux distribution in $B_x$ for 11 energies values ranging from 0.49 keV up to 29.0 keV. (b) THEMIS-E measured parallel electron energy flux distribution in $B_x$ in the same format as (a). (c) Statistical distribution of equatorial energy spectra, along with the medium and maximum energy flux levels, measured by THEMIS-D and THEMIS-E when $B_x$ was within 10--35 nT near the equator. }
\label{figS2}
\end{figure}
\clearpage
\newpage

\begin{figure}
\setfigurenum{S2}
\centering
\includegraphics[scale=0.85,angle=0,origin=c]{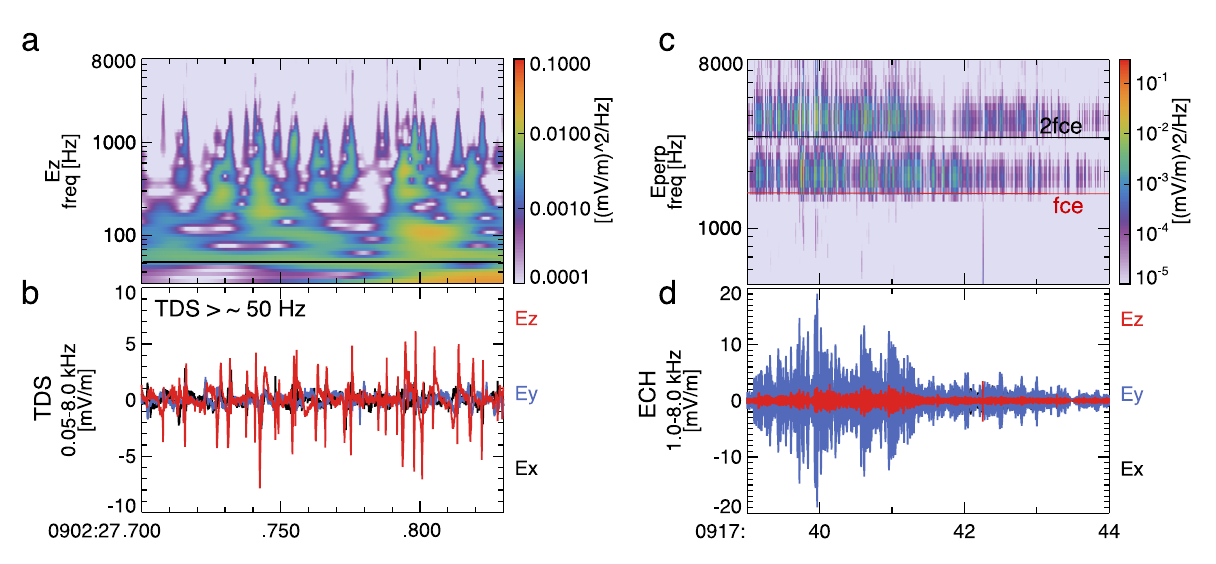}
\end{figure}
\newpage
\begin{figure}
\caption{Examples of THEMIS-E waveform measurements showing TDSs within broadband electrostatic fluctuations and ECH waves around injections and braking ion flows. (a-b) Wavelet spectrogram of $E_z$ and waveform time series of TDS electric fields in the field-aligned coordinates, measured during 09:02:27.7--09:02:27.83 UT. (c-d) Wavelet spectrogram of $E_{\perp}$ and waveform time series of ECH wave electric fields, measured during 09:17:39--09:17:44 UT.}
\label{figS3}
\end{figure}
